\documentclass[prd,twocolumn,a4paper,superscriptaddress,floatfix]{revtex4}
\usepackage{graphicx}
\usepackage{bbm}
\usepackage[all]{xy}
\usepackage{amsmath}
\usepackage{amssymb}
\usepackage{epstopdf}

\def\be{\begin{equation}}
\def\ee{\end{equation}}
\newcommand{\bq}{\begin{eqnarray}}
\newcommand{\eq}{\end{eqnarray}}
\newcommand{\bes}{\begin{subequations}}
\newcommand{\ees}{\end{subequations}}

\def\ben{\begin{eqnarray}}
\def\een{\end{eqnarray}}
\def\ba{\begin{array}}
\def\ea{\end{array}}

\begin{document}
\newcommand{\half}{{\textstyle\frac{1}{2}}}
\allowdisplaybreaks[3]
\def\a{\alpha}
\def\b{\beta}
\def\g{\gamma}\def\G{\Gamma}
\def\d{\delta}\def\D{\Delta}
\def\ep{\epsilon}
\def\et{\eta}
\def\z{\zeta}
\def\t{\theta}\def\T{\Theta}
\def\l{\lambda}\def\L{\Lambda}
\def\m{\mu}
\def\f{\phi}\def\F{\Phi}
\def\n{\nu}
\def\p{\psi}\def\P{\Psi}
\def\r{\rho}
\def\s{\sigma}\def\S{\Sigma}
\def\ta{\tau}
\def\x{\chi}
\def\o{\omega}\def\O{\Omega}
\def\k{\kappa}
\def\pa {\partial}
\def\ov{\over}
\def\br{\\}
\def\ud{\underline}

\newcommand\lsim{\mathrel{\rlap{\lower4pt\hbox{\hskip1pt$\sim$}}
    \raise1pt\hbox{$<$}}}
\newcommand\gsim{\mathrel{\rlap{\lower4pt\hbox{\hskip1pt$\sim$}}
    \raise1pt\hbox{$>$}}}
\newcommand\esim{\mathrel{\rlap{\raise2pt\hbox{\hskip0pt$\sim$}}
    \lower1pt\hbox{$-$}}}

\title{$\chi^{2}$ versus median statistics in SNIa data analysis}

\author{A. Barreira}
\email[Electronic address: ]{alex.mr.barreira@hotmail.com}
\affiliation{Centro de F\'{\i}sica do Porto, Rua do Campo Alegre 687, 4169-007 Porto, Portugal}
\affiliation{Departamento de F\'{\i}sica da Faculdade de Ci\^encias
da Universidade do Porto, Rua do Campo Alegre 687, 4169-007 Porto, Portugal}
\author{P.P. Avelino}
\email[Electronic address: ]{ppavelin@fc.up.pt}
\affiliation{Centro de Astrof\'{\i}sica da Universidade do Porto, Rua das Estrelas, 4150-762 Porto, Portugal}
\affiliation{Departamento de F\'{\i}sica da Faculdade de Ci\^encias
da Universidade do Porto, Rua do Campo Alegre 687, 4169-007 Porto, Portugal}

\begin{abstract}
In this paper we compare the performances of the $\chi^{2}$ and median likelihood analysis in the determination of cosmological constraints using type Ia supernovae data. We perform a statistical analysis using the 307 supernovae of the Union 2 compilation of the Supernova Cosmology Project and find that the $\chi^{2}$ statistical analysis yields tighter cosmological constraints than the median statistic if only supernovae data is taken into account. We also show that when additional measurements from the Cosmic Microwave Background and Baryonic Acoustic Oscillations are considered, the combined cosmological constraints are not strongly dependent on whether one applies the $\chi^{2}$ statistic or the median statistic to the supernovae data. This indicates that, when complementary information from other cosmological probes is taken into account, the performances of the $\chi^2$ and median statistics are very similar, demonstrating the robustness of the statistical analysis.
\end{abstract} 
\maketitle

\section{Introduction}

More than a decade ago, type Ia supernovae (SNIa) provided the first clear evidence in favor of cosmic acceleration \cite{Riess:1998cb,Perlmutter:1998np}. Since then, the availability of ever larger, higher-quality SNIa datasets, as well as measurements using other cosmological probes, such as the Cosmic Microwave Background (CMB) or the Baryonic Acoustic
Oscillations (BAO), have been providing overwhelming evidence for the existence of dark energy \cite{Frieman:2008sn,Amanullah:2010vv,Komatsu:2010fb}, a fluid with large negative pressure capable of driving the acceleration of the universe.

In the so-called $\Lambda{\rm CDM}$ model, also known as the concordance model, the dark energy role is played by a cosmological constant $\Lambda$, responsible for approximately 73\% of the energy density of the universe at the present day. The remaining percentage is mainly in the form of cold matter, most of which non-baryonic and dark. Radiation is residual at the present time and the universe is spatially flat. Despite the good agreement with observational data, this model has little appeal on theoretical grounds since the value of $\Lambda$ required to explain the observed cosmic acceleration is off by $\sim120$ orders of magnitude from the standard quantum field theory prediction. The coincidence between our observing time and the time of the onset of cosmic acceleration is also puzzling \cite{Avelino:2004vy,Barreira:2011qi}. Many other dark energy models have been proposed, with the most influential being the ones based on dynamical scalar fields. In these models, the energy density varies with time and suitable choices of the scalar field lagrangian can relax some of the problems associated with the cosmological constant (see, for instance, the review \cite{Copeland:2006wr}). As a result, an important step towards a better understanding of the dark energy involves further testing of its possible dynamical nature.

The determination of dark energy constraints from observational data requires a robust statistical data analysis. In the particular case of SNIa, the usual procedure is to carry out a standard $\chi^{2}$ likelihood analysis. However, in \cite{Gott:2000mv,Avelino:2001xj} it has been argued that the median statistic could be a more reliable alternative. The use of the median, despite having the drawback of not being as contraining as the $\chi^{2}$ analysis, has the strong advantage of requiring weaker assumptions about the data, thus yielding more trustworthy constraints. Moreover, the median is less vulnerable to the presence of {}``outliers'', which is also a significant advantage given current uncertainties about the physics of SNIa.

In this work we revisit and extend the analysis of \cite{Avelino:2001xj}, updating it using the recent Union 2 SNIa compilation of the Supernova Cosmology Project (SCP) \cite{Amanullah:2010vv}. We compare the performances of $\chi^{2}$ and median statistics in the determination of cosmological constraints using SNIa data, considering also CMB and BAO measurements. The layout of this paper is as follows. In Sec. II we describe the application of $\chi^{2}$ and median statistics to SNIa data. We also discuss the use of additional information contained in the CMB and BAO. In Sec. III we present and discuss our results. Finally, we conclude in Sec. IV.

\section{Cosmological probes}

\subsection{SNIa}

Type Ia supernovae appear to be good {}``standard candles'' and therefore they can serve as useful distance indicators. Constraints arise by comparing the theoretical distance modulus,
\begin{equation}
\mu_{th}=5\log_{10}\left(\frac{d_{L}(z)}{{\rm Mpc}}\right)+25\,,\label{eq:theoretical distance modulus}
\end{equation}
with the observational distance modulus $\mu_{obs}$ inferred from the data, at the measured supernovae redshifts. In Eq. (\ref{eq:theoretical distance modulus}), $d_{L}(z)$ is the luminosity distance given by,
\begin{equation}
d_{L}(z)=\frac{(1+z)c}{H_{0}\sqrt{|\Omega_{k0}|}}S\left(\sqrt{|\Omega_{k0}|}\int_{0}^{z}\frac{dz'}{E(z')}\right)\,,\label{eq:luminosity distance}
\end{equation}
where $z$ is the cosmological redshift, $H_{0}$ is the present day value of the Hubble expansion rate and $E(z)=H(z)/H_{0}$. The function $S$ is defined as,
\begin{equation}
S(x)=\begin{cases}
\begin{array}{c}
\sin x,\\
x,\\
\sinh x,\end{array} & \begin{array}{c}
\Omega_{k0}<0\,,\\
\Omega_{k0}=0\,,\\
\Omega_{k0}>0\,,\end{array}\end{cases},
\end{equation}
with the flat case ($\Omega_{k0}=0$) being recovered taking the limit $\Omega_{k0}\rightarrow0$ in Eq. (\ref{eq:luminosity distance}). In this paper we consider, 
\begin{equation}
E(a)=\sqrt{\Omega_{m0}a^{-3}+\Omega_{de0}a^{-3(1+w)}+\Omega_{k0}a^{-2}}\,,\label{eq: cosmological model}
\end{equation}
where $a=1/(1+z)$ is the scale factor, $\Omega_{m0}$ and $\Omega_{de0}$ are, respectively, the present day values of the matter and dark energy fractional densities ($\Omega_{i}=\rho_{i}/\rho_{c}$, with $\rho_{c}$ being the critical density) and $\Omega_{k0}=1-\Omega_{m0}-\Omega_{de0}$. The parameter $w$ is the equation of state parameter of the dark energy (the ratio between the pressure and the energy density) which, for simplicity, we assume to be constant (see however \cite{Avelino:2009ze,Avelino:2011ey}). If the dark energy role is played by a cosmological constant $\Lambda$ then $\Omega_{de}=\Omega_\Lambda$ and $w=-1$.

\subsubsection{$\chi^{2}$ analysis}

In the $\chi^{2}$ statistical analysis, the likelihood $P$ of the cosmological parameters $Q$ is given by $P\propto\exp\left(-\chi^{2}/2\right)$, with 
\begin{equation}
\chi^{2}=\sum_{i=1}^{N}\left(\frac{\mu_{obs}(z_{i})-\mu_{th}(z_{i},Q)}{\sigma_{i}}\right)^{2}\,.\label{eq:chi^2}
\end{equation}
In Eq. (\ref{eq:chi^2}), $N$ is the number of supernovae in the dataset and $\sigma_{i}$ is the observational error associated with
$\mu_{obs}$ at the redshift $z_{i}$. The $\chi^2$ analysis assumes that: (i) the experimental results are statistically independent; (ii) there are no systematic errors; (iii) the statistical errors follow a Gaussian distribution; (iv) the standard deviation of the statistical errors is equal to the observational uncertainty. 

Presently, there is no strong evidence supporting that the supernovae magnitude errors are Gaussianly distributed and therefore the hypothesis (iii) and (iv) are quite strong. Moreover, the $\chi^{2}$ statistic is highly susceptible to the presence of {}``outliers'' in the datasets. This constitutes an extra concern, in particular due to the uncertainties associated to the calibration of the SNIa light curves.

\subsubsection{Median analysis}

The fewer the assumptions one needs to make about the dataset, the more reliable the results derived from it are. The median statistical analysis discards the assumption of the Gaussianity of the errors, requiring only the use of hypothesis (i) and (ii). 

Assuming that the experimental results are statistically independent and that there are no systematic errors present, one expects that after performing a sufficiently large number of measurements, approximately, half of the values obtained will be above (or below) the correct mean value. In particular, if we perform $N$ measurements, the probability that $k$ of them will be above (or below) the median is given by the binomial distribution,
\begin{equation}
P(k,N)=\frac{2^{-N}N!}{k!(N-k)!}\,.\label{eq:binomical distribution}
\end{equation}
This way, given the data set $\mu_{obs}(z_{i})\ (i=1..N)$, the likelihood of the cosmological parameters $Q$ is obtained by counting the number of observational values that fall above (or below) the curve given by Eq. (\ref{eq:theoretical distance modulus}).

Despite being associated with less tighter constraints than the $\chi^2$ statistic, the median statistic is not very sensitive to the presence of a few {}``outlier'' SNIa objects. In \cite{Gott:2000mv}, it has been shown how the presence of one or very few {}``ill'' data points could severely distort a $\chi^{2}$ analysis, while the median results remained approximately the same.

On the other hand, one should notice that the likelihood computed with the median statistic only accounts for the number of experimental points above or below the expected correct curve, not differentiating between the various ways in which these points could be distributed. For instance, a set where the first half of the data points is above the expected curve and the second half is below has the same likelihood (when computed with the median) as a set where the first point is above, the second below, the third above, the fourth below, and so forth. These two cases should not be indistinct since the first could turn out to be a terrible fit to the data. In \cite{Avelino:2001xj} it has been shown how some modifications of the median statistic could alleviate this problem. In short, the modifications involved taking into account the size of the largest continuous sequence found above (or below) the model's prediction, or the number of total continuous sequences obtained.

Here, we propose an alternative way to cope with this problem. Instead of counting the number of points that are above (or below) the model's prediction in the entire redshift range of the data set, one may divide the dataset into redshift intervals with a certain number of SNIa objects and perform the counting in each interval. Suppose we divide the dataset into $n$ intervals, with $N_{j}$ being the number of SNIa objects in the $j$-th interval ($j=1..n)$. This way, the overall likelihood of the parameters $Q$ is given by
\begin{equation}
P=\prod_{j=1}^{n}P(k_{j},N_{j})\,,
\end{equation}
where $k_{j}$ is the number of points that, in the $j$-th interval, are above (or below) the theoretical curve given by the parameters. This way, by properly dividing the dataset into groups of supernovae, we are more likely to avoid pathological situations in which very large sequences of SNIa above (or below) the median value are present.

\begin{figure}
\includegraphics[width=8cm]{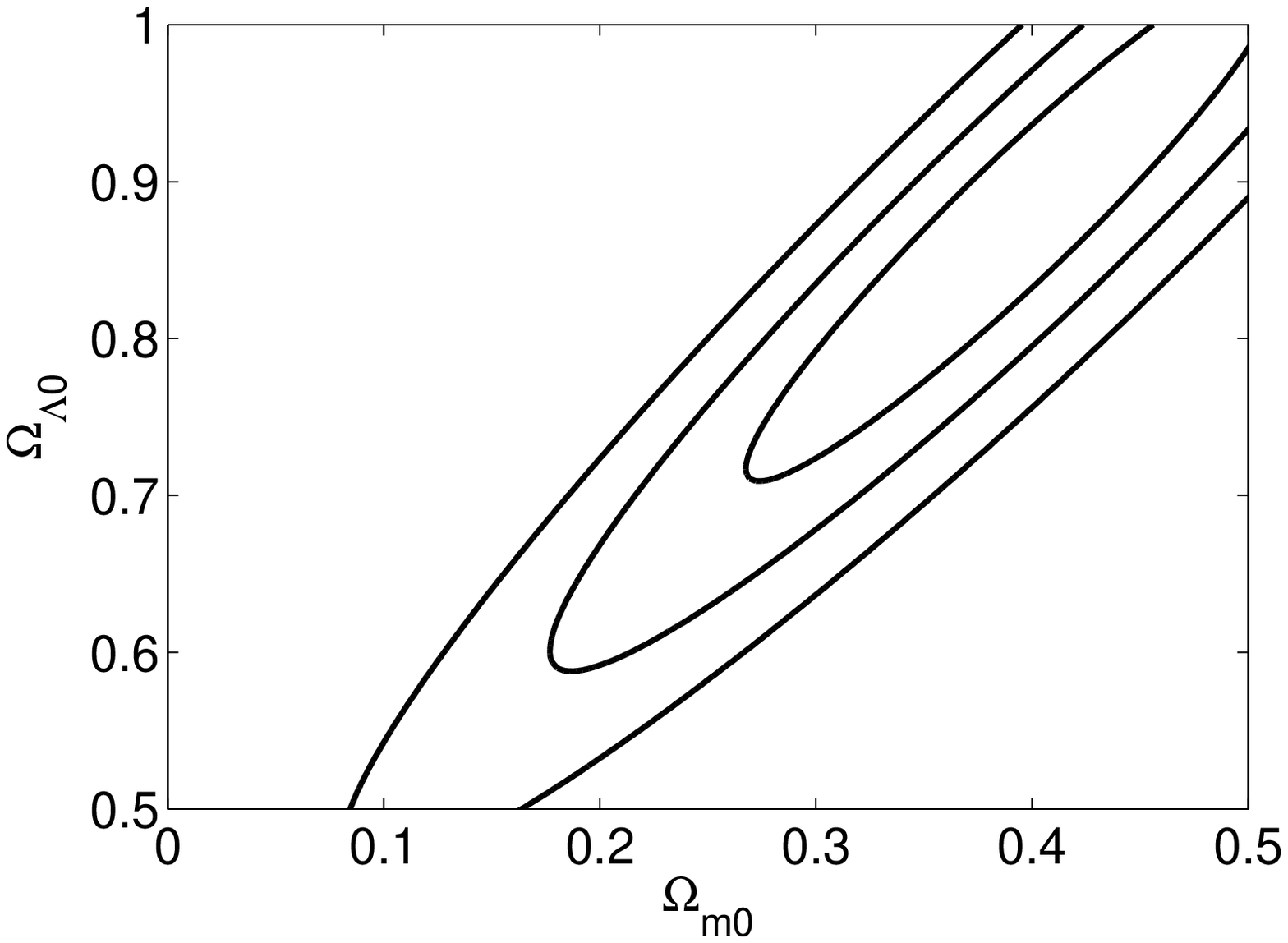}\caption{\label{fig:1} $68.3\%,\ 95.4\%\  {\rm and\ }99.7\%$ confidence level
contours on $\Omega_{m0}$ and $\Omega_{\Lambda 0}$, obtained from the Union 2 SNIa dataset using the $\chi^{2}$ statistical analysis.}
\end{figure}

\begin{figure}
\includegraphics[width=8cm]{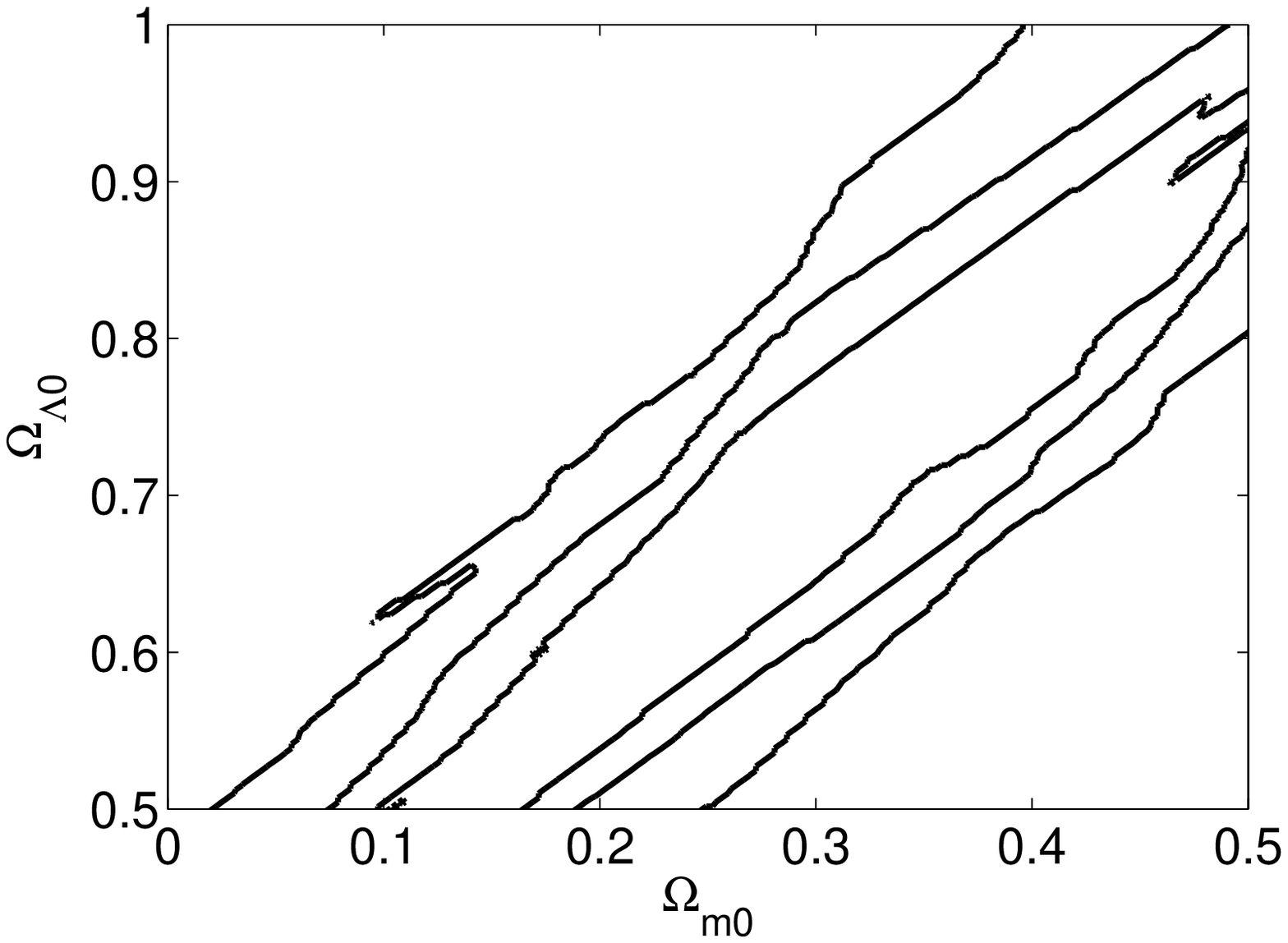}\caption{\label{fig:2} Same as Fig. \ref{fig:1} but using the median statistical analysis.}
\end{figure}

\begin{figure}
\includegraphics[width=8cm]{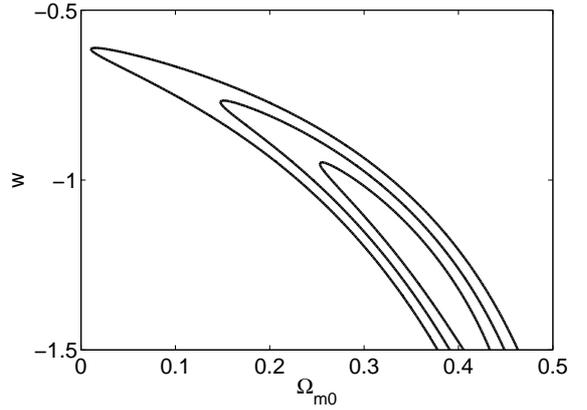}\caption{\label{fig:3} $68.3\%,\ 95.4\%\  {\rm and\ }99.7\%$ confidence level contours on $\Omega_{m0}$ and $w$ (assuming that $\Omega_{k0}=0)$, obtained from the Union 2 SNIa dataset using the $\chi^{2}$ statistical analysis.}
\end{figure}

\begin{figure}
\includegraphics[width=8cm]{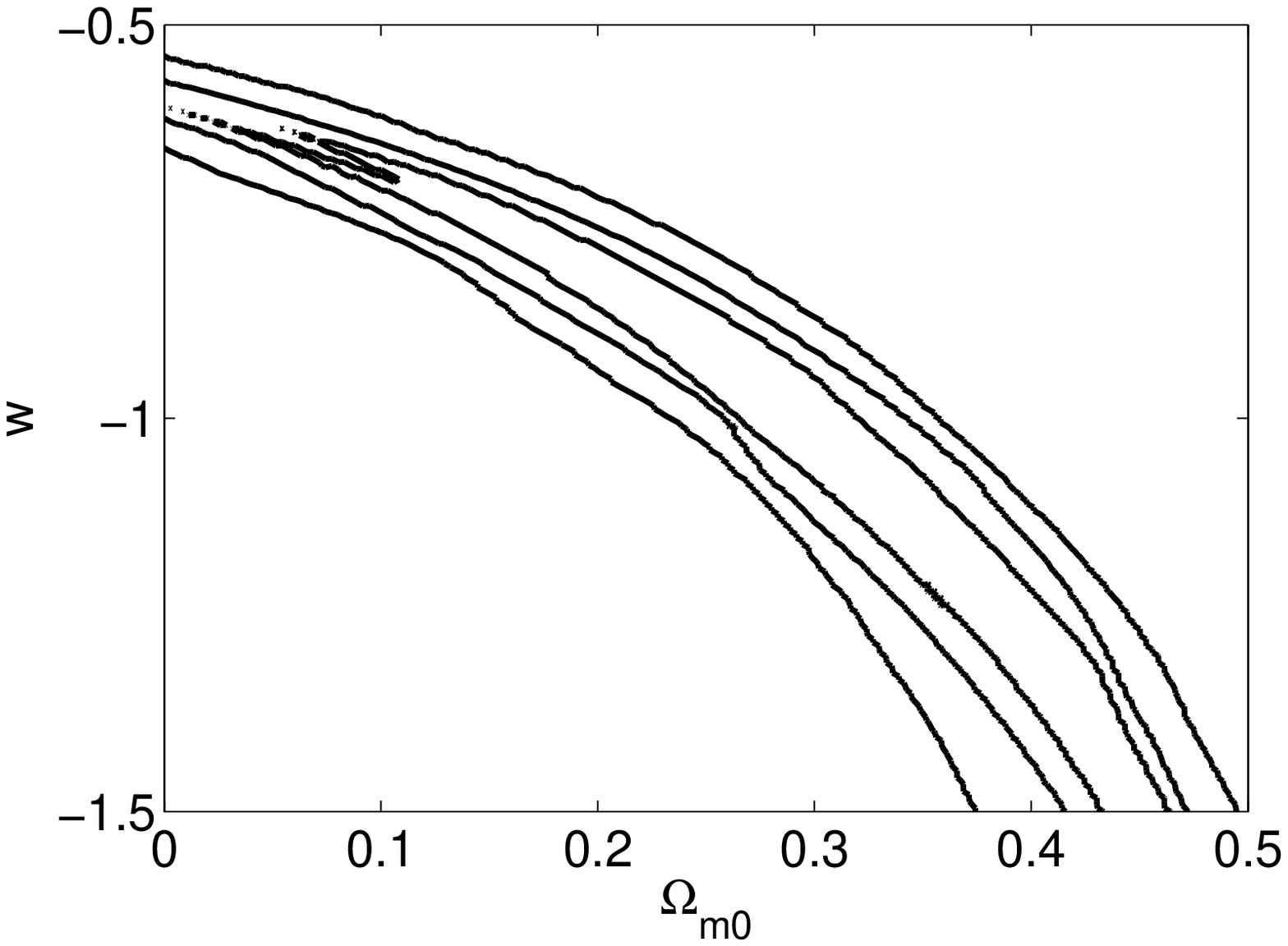}\caption{\label{fig:4} Same as Fig. \ref{fig:3} but using the median statistical analysis.}
\end{figure}

\subsection{The CMB shift parameter}

The CMB shift parameter $R$ is defined by
\begin{eqnarray}
R_{th}&\equiv&\sqrt{\Omega_{m0}}\frac{H_{0}}{c}\frac{d_{L}(z_{dec})}{(1+z_{dec})}=\nonumber\\
&=&\frac{\sqrt{\Omega_{m0}}}{\sqrt{|\Omega_{k0}|}}S\left(\sqrt{|\Omega_{k0}|}\int_{0}^{z_{dec}}\frac{dz'}{E(z')}\right)\,,\label{eq:CMB shift parameter}
\end{eqnarray}
with the redshift of decoupling $z_{dec}\approx1090.97$ \cite{Hu:1995en,Komatsu:2010fb}. Following the WMAP 7-year results \cite{Komatsu:2010fb} we take $R_{obs}=1.725$ with an error $\sigma_{R}=0.018$. The likelihood derived from the shift parameter is then $P_{CMB}\propto\exp\left(-\chi_{CMB}^{2}/2\right)$, with
\begin{equation}
\chi_{CMB}^{2}=\left(\frac{R_{th}-R_{obs}}{\sigma_{R}}\right)^{2}\,.\label{eq:chi2 CMB}
\end{equation}
It turns out that assuming that $R$ is Gaussianly distributed around $R_{obs}$ with standard deviation $\sigma_{R}$ provides an efficient summary of the information encoded in the full CMB data \cite{Wang:2007mza,Elgaroy:2007bv}.

\subsection{The BAO scale}

\begin{figure}
\includegraphics[width=8cm]{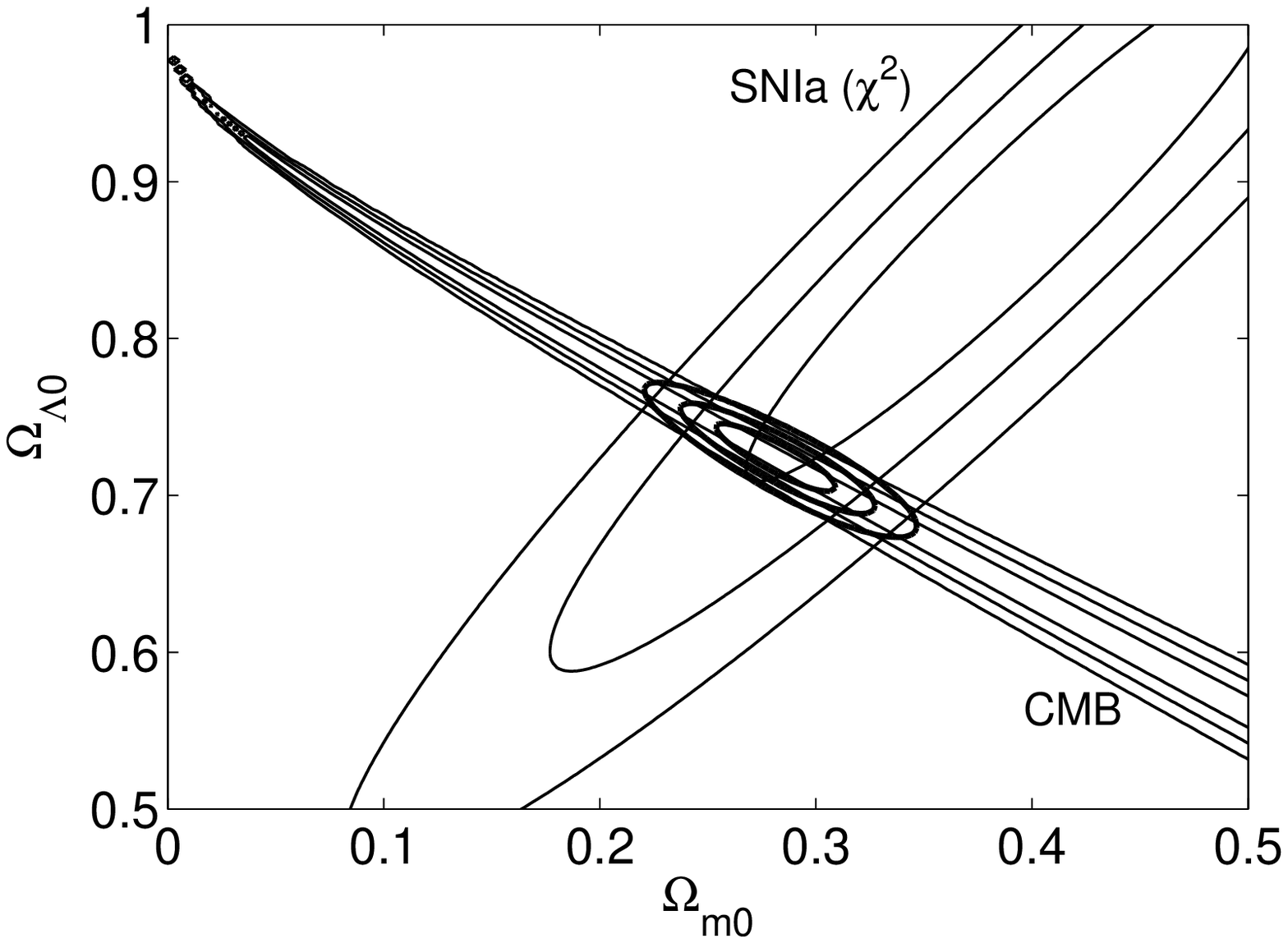}\caption{\label{fig:5} $68.3\%,\ 95.4\%\  {\rm and\ }99.7\%$ confidence level contours on $\Omega_{m0}$ and $\Omega_{\Lambda 0}$, obtained from the Union 2 SNIa dataset, combined with constraints on the CMB shift parameter $R$, using the $\chi^{2}$ statistical analysis.}
\end{figure}

\begin{figure}
\includegraphics[width=8cm]{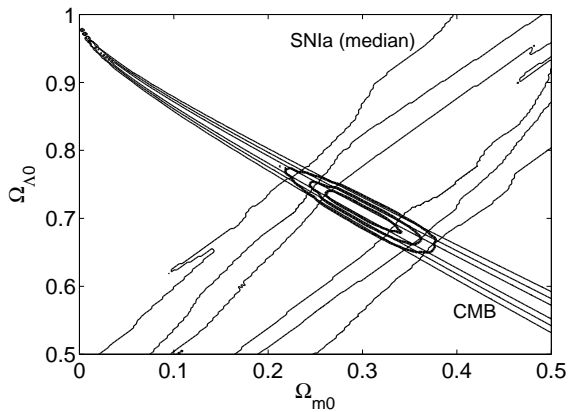}\caption{\label{fig:6} Same as Fig. \ref{fig:5} but using the median statistical analysis for the SNIa data.}
\end{figure}

The baryonic acoustic oscillations imprinted in the CMB manifest themselves today in the large-scale distribution of galaxies. The BAO signature on large scales was found in \cite{Eisenstein:2005su}, when a small {}``bump'' in the two-point correlation function of red-luminous galaxies was measured. Cosmological constraints arise via the position of the {}``bump'', which is related to the quantity
\begin{eqnarray}
A_{th}(z)&=&\sqrt{\Omega_{m0}}E(z)^{-\frac13} \times \nonumber\\
&\times& \left[\frac{1}{z\sqrt{|\Omega_{k0}|}}S\left(\sqrt{|\Omega_{k0}|}\int_{0}^{z}\frac{dz'}{E(z')}\right)\right]^{\frac23}\,.\label{eq:BAO A}
\end{eqnarray}
The likelihood is $P_{A}\propto\exp\left(-\chi_{A}^{2}/2\right)$, with
\begin{equation}
\chi_{A}^{2}=\left(\frac{A_{th}(z_{BAO})-A_{obs}}{\sigma_{A}}\right)^{2}\,,\label{eq:chi2 A}
\end{equation}
where $z_{BAO}=0.35$, $A_{obs}=0.469$ and $\sigma_{A}=0.017$ \cite{Percival:2009xn,Reid:2009xm}. Just like in the case of the shift parameter, it is usually safe to assume that $A$ is Gaussianly distributed \cite{Eisenstein:2005su,Percival:2009xn}.

\begin{figure}
\includegraphics[width=8cm]{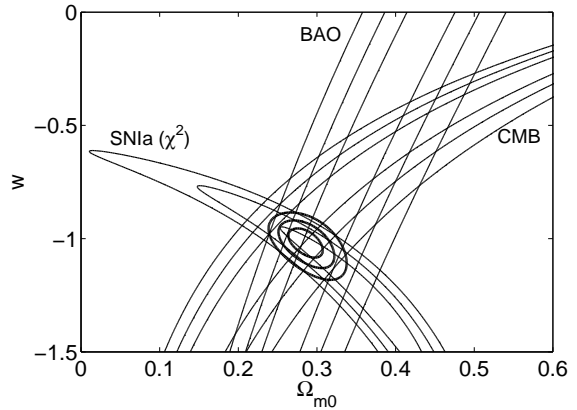}\caption{\label{fig:7}  $68.3\%,\ 95.4\%\  {\rm and\ }99.7\%$ confidence level contours on $\Omega_{m0}$ and $w$ (assuming that $\Omega_{k0}=0)$, obtained from the Union 2 SNIa dataset, combined with constraints on the CMB shift parameter $R$ and the BAO scale, using the $\chi^{2}$ statistical analysis.}
\end{figure}

\begin{figure}
\includegraphics[width=8cm]{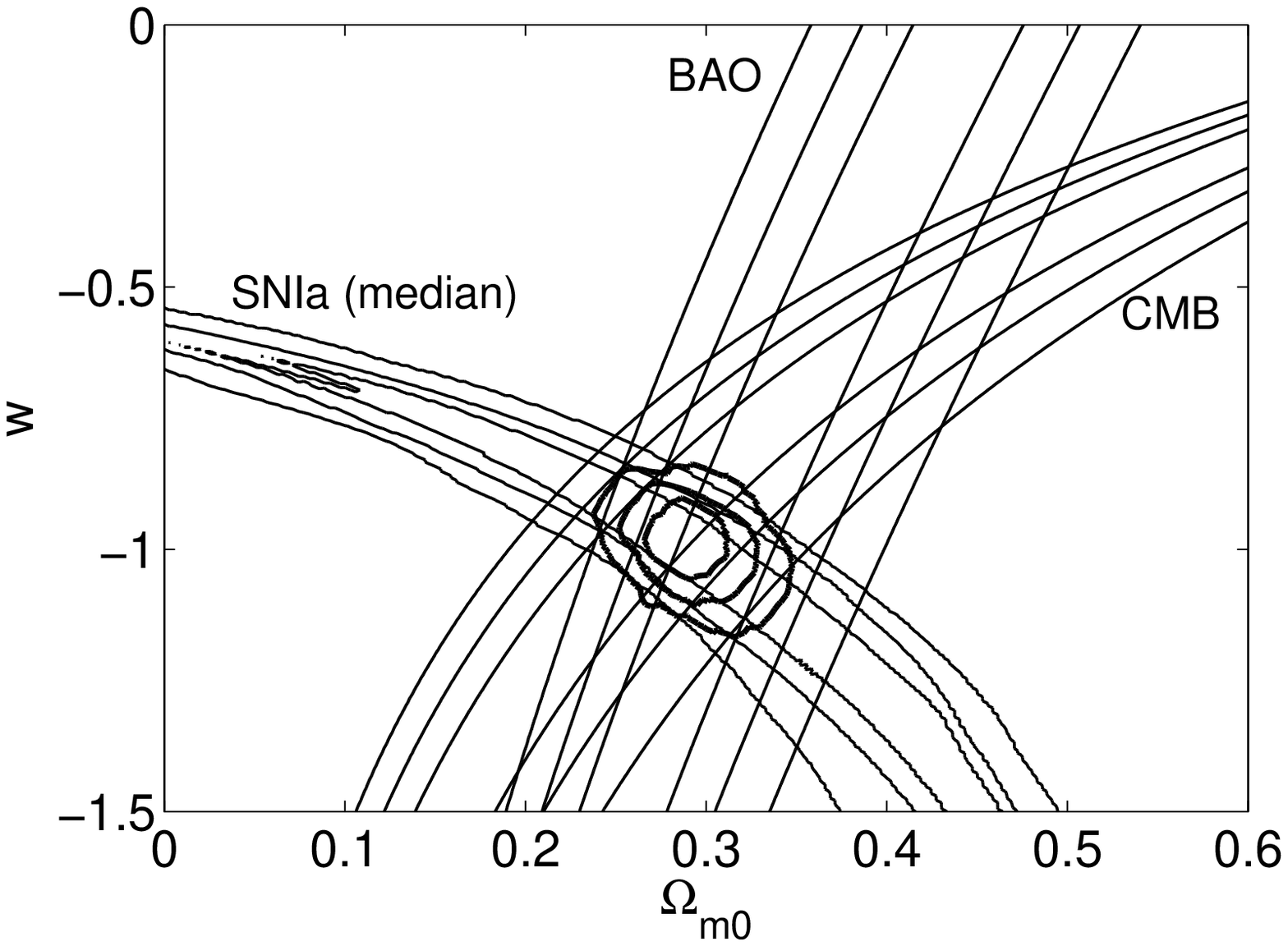}\caption{\label{fig:8} The same as Fig. \ref{fig:7} but using the median statistical analysis for the SNIa data.}
\end{figure}

\begin{table*}
\begin{tabular}{|c|c|c|c|}
\hline 
Fit & $\Omega_{m0}$ & $\Omega_{k0}$ & $w$\tabularnewline
\hline
\hline 
SNIa ($\chi^{2}$) + CMB & $\left[0.254;0.310\right]$ & $\left[-0.043;0.056\right]$ & $-1({\rm fixed)}$\tabularnewline
\hline 
SNIa (median) + CMB & $\left[0.260;0.342\right]$ & $\left[-0.061;0.083\right]$ & $-1({\rm fixed)}$\tabularnewline
\hline 
SNIa ($\chi^{2}$) + CMB + BAO & $\left[0.269;0.313\right]$ & $\left[-0.031;0.049\right]$ & $-1({\rm fixed)}$\tabularnewline
\hline 
SNIa (median) + CMB + BAO & $\left[0.275;0.333\right]$ & $\left[-0.042;0.064\right]$ & $-1({\rm fixed)}$\tabularnewline
\hline 
SNIa ($\chi^{2}$) + CMB + BAO & $\left[0.264;0.308\right]$ & $0({\rm fixed)}$ & $\left[-1.084;-0.955\right]$\tabularnewline
\hline 
SNIa (median) + CMB + BAO & $\left[0.266;0.312\right]$ & $0({\rm fixed)}$ & $\left[-1.058;-0.902\right]$\tabularnewline
\hline
\end{tabular}
\caption{\label{tab:1} Constraints on $\Omega_{m0}$, $\Omega_{de0}$ and $w$, obtained from SNIa ($\chi^{2}$ and median analysis), CMB and BAO data, as well as their combinations. The constraints are at the 1$\sigma$ level.}
\end{table*}

\section{Results}

\subsection{SNIa constraints}

Following the discussion in Sec. IIA, we divide the SNIa Union 2 dataset \cite{Amanullah:2010vv}, which containts a total of $N=307$ SNIa objects, in groups of approximately 75 SNIa each. The supernovae are ordered by increasing redshift  with the first three groups having 76 supernovae and the forth group 79. Several other ways to divide the set are possible. However, this choice is not critical for our conclusions.

Figures \ref{fig:1} and \ref{fig:2} show the $68.3\%,\ 95.4\%\  {\rm and\ }99.7\%$ confidence level contours on $\Omega_{m0}$ and $\Omega_{\Lambda 0}$, obtained from the Union 2 SNIa dataset using the $\chi^{2}$ and median statistical analysis, respectively.
Figures \ref{fig:3} and \ref{fig:4} are similar to Figs. \ref{fig:1} and \ref{fig:2}, except that now the SNIa constraints are on the ($\Omega_{m0}$,$w$) plane and the condition $\Omega_{k0}=0$ was assumed. In both cases the Hubble parameter today was held fixed at $H_{0}=70.2 \, {\rm km/s/Mpc}$ \cite{Komatsu:2010fb}. As expected, the constraints using the $\chi^{2}$ statistic are tighter, consequence of the stronger assumptions it makes about the data. Nevertheless, the constraints obtained using the median statistical analysis are not as bad as one could originally fear. The median contours are more {}``stretched'' than the $\chi^2$ ones, but their {}``width'' is similar. The median statistical constraints are in principle more reliable, since they do not assume Gaussianity of the SNIa magnitude error distribution.

\subsection{SNIa + CMB + BAO constraints}

Figures \ref{fig:5} and \ref{fig:6} show the $68.3\%,\ 95.4\%\  {\rm and\ }99.7\%$ confidence level contours on $\Omega_{m0}$ and $\Omega_{\Lambda 0}$, obtained from the Union 2 SNIa dataset (using the $\chi^{2}$ and median statistical analysis, respectively), combined with constraints on the CMB shift parameter $R$. The differences between the results obtained with the $\chi^{2}$ and median statistics are now significantly reduced. This is related to the fact that, as shown in Figs.  \ref{fig:1} and \ref{fig:2}, the {}``width'' of the SNIa contours, obtained using the $\chi^{2}$ and median analysis, is very similar. The CMB contours {}``cross'' the SNIa contours, rendering similar combined constraints in both cases. Considering the BAO data (not shown) tightens the constraints slightly (see Table 1).

Figures \ref{fig:7} and \ref{fig:8} show the combined constraints on $\Omega_{m0}$ and $w$ (assuming that $\Omega_{k0}=0$), obtained from SNIa (using $\chi^{2}$ and median analysis, respectively), CMB and BAO data. Again, we see that the combined constraints are weakly dependent on whether one derives the constraints from SNIa with the $\chi^{2}$ or with the median statistical analysis.

Table \ref{tab:1} presents a summary of the constraints, at the $1\sigma$ level, obtained from different data combinations and
model assumptions. It shows that the median is almost as constraining as the $\chi^{2}$ statistical analysis, if combined with additional CMB and BAO constraints. Hence, the strong assumption that the SNIa measurements are Gaussianly distributed is not necessary in order to obtain tight constraints. On the other hand, since the $\chi^{2}$ and median statistical analysis yield similar cosmological constraints when the SNIa data is combined with other cosmological probes, our results may also be interpreted as providing additional validation of the results obtained using the standard $\chi^{2}$ statistical analysis.

\section{Conclusion}

In this paper we compared the cosmological constraints derived from the Union 2 SNIa dataset using $\chi^{2}$ and median statistics. In the absence of CMB and BAO constraints, we have shown that the $\chi^{2}$ statistic yields tighter cosmological constraints than the median statistic, as a result of the stronger assumptions it makes about the SNIa magnitude error distribution. On the other hand, when CMB and BAO information is taken into account, the performances of both statistics are very similar. Hence, we conclude that the assumption of the gaussian distribution of the errors in the SNIa analysis does not appear to be critical in the determination of cosmological constraints, provided that complementary information from other cosmological probes is also taken into account.


\bibliography{median}

\end{document}